\documentstyle[aps,prd,twocolumn]{revtex}

\begin{document}
\draft

\title{THE INTERIOR FIELD OF A MAGNETIZED \\ EINSTEIN-MAXWELL OBJECT}
\author
{Tonatiuh Matos\thanks{E-mail: tmatos@fis.cinvestav.mx},\\ 
Departamento de F\'{\i}sica,\\ Centro de Investigaci\'on y de Estudios
Avanzados del IPN,\\ P.O. Box 14-740, M\'exico 07000, D.F., M\'exico.}

\date{ \today}
%
%%%%% Remove the next line for preprint style:
\twocolumn[
\maketitle

\begin{abstract}
Using the Harmonic map ansatz, we reduce the axisymmetric, static
Einstein-Maxwell equations coupled with a magnetized perfect fluid to a set
of Poisson-like equations. We were able to integrate the Poisson equations
in terms of an arbitrary function $M=M(\rho,\zeta)$ and some integration
constants. The thermodynamic equation
restricts the solutions to only some  state equations, but in some cases
when the solution exists, the interior solution can be matched with the
corresponding exterior one.
\end{abstract}


%
%%%%% Remove the next line for preprint style:
]
\narrowtext

\newpage

Exact solutions of the Einstein's field equations have been one of the most
interesting challenge for mathematicians and physicist \cite{kra}. A great
effort has been done for finding exact solutions with a physical
interpretation. In the seventies mathematicians and relativists had a great
success using very elaborate mathematical technics, a great amount of exact
solutions of the Einstein equations have been therefore found, analyzed and
studied in the seventies and eighties. The discovery of the binary
pulsars gave rise to non-perturbative effects of general relativity and
the exact solutions of the 
Einstein's field equations become a necessarily subject for astrophysicists
and relativists. Furthermore, for compact objects like white dwarfs,
pulsars and black holes, non-pertubative effects are the most interesting
one, effects which can only be understood better with exact solutions of
the field equations.

In this letter I am interested in the interior field of an object, taking
matter into account. I will suppose axial symmetry only, and allow metric
functions with as most freedom as we can. As a first approximation I will
suppose that the metric is static. Now we must determine the right hand
side of the Einstein equations. To describe this object one can start
modeling its interior matter as a perfect fluid. It is difficult to know
the state equation of a compact object, we do not use to work with matter
 in so extreme conditions of density and pressure. We even do not know if
the energy conditions for the energy momentum tensor are valid at this
extreme. Something we can do, is to make a general ansatz about the state
equation, for example $p=\omega\mu$,
 where $\omega$ is a constant, $p$ is the pressure of matter and $\mu$ is
its energy density, and let that the theory says us something about this
ansatz. As a first approximation this ansatz seems to be reasonable, but to
have a more complete study of this class of matter, we must investigate
other possibilities. 

Let me start with this ansatz and investigate what the Einstein-Maxwell
theory can say about it.
The action we deal with is then
\begin{eqnarray}
{\cal S} &= \int d^4x\sqrt{-g} [ - R +F^{\mu\nu}F_{\mu\nu} + 
{\cal L}_{Matter}]
\label{action}
\end{eqnarray}
where $g=\det(g_{\mu\nu}), \ \mu, \nu = 0,1,2,3$, $R$ is the scalar
curvature and  $F_{\mu\nu}$ is the Maxwell tensor. 

After variation with respect to the electromagnetic field and the metric,
we respectively obtain the following field equations
\begin{eqnarray}
&F^{\mu\nu}_{\ \ \ \ ;\nu} = 0, \cr  \cr
&G_{\mu\nu} =  2 (F_{\mu\lambda}F_\nu^{\ \lambda} - {1\over 4} g_{\mu\nu}
F_{\alpha\beta}F^{\alpha\beta} ) + 2 T_{pf\mu\nu}\ , \cr\cr
&T_{pf\ \ \ \ ;\nu}^{\ \ \mu\nu}=0
\label{eq}
\end{eqnarray}
where the last field equation in (\ref{eq}) is the thermodynamic one. In
this work we will only consider the axisymmetric static case, that means,
space-times containing one space-like and one time-like commuting,
hypersurface forming Killing vector, and 
the case when matter corresponds to a perfect fluid, $i.e.\ \
{T_{pf}}^{\mu}_{\ \nu}=$diag$(p,p,p,-\mu)$. The axisymmetric static metric
convenient for the energy conditions of this case is the static Papapetrou
metric \cite{kra}
\begin{equation}
ds^2={1\over f}\{e^{2k}\ (d\rho^2+ d\zeta^2) + W^2 d\varphi^2\}-f\ dt^2
\label{papa}
\end{equation}
where $f,\ W,$ and $k$ are functions of $\rho$ and $\zeta$ only. Now we use
the harmonic map ansatz. Suppose that $\lambda$ and $\tau$ are coordinates
of a two dimensional flat space $ds^2_{v_p}=d\lambda\ d\tau,$ with
$\lambda=\lambda(\rho,\zeta)$ and $\tau=\tau(\rho,\zeta)$ and suppose that
$f=f(\lambda,\ \tau)$. Using the procedure of reference \cite{ma25,ma36} we
find that $f=e^{\lambda-\gamma\tau}$ is a good parametrization for a
space-time curved by a gravitational and an electromagnetic field. In terms
of the functions $\lambda$ and $\tau$ the axisymmetric static Einstein
field equations for the action (\ref{action}) read
\begin{eqnarray}
2\Delta\lambda &= &(3p+\mu)\sqrt{-g}\cr\cr
\Delta\ln W &= &2p\sqrt{-g}\cr\cr
\Delta\tau +2\ W((\tau_{,\rho})^2&+
&(\tau_{,\zeta})^2)e^{\lambda-(\gamma-2)\tau}=0 
\label{poisson}
\end{eqnarray}
where $\Delta f={1\over 2}[(Wf_{,\rho})_{,\rho}+ (Wf_{,\zeta})_{,\zeta}]$
is the generalized Laplace operator. The electromagnetic four potential is
given as $(A_\rho,A_{\zeta},A_{\varphi},A_t)=(0,0,A_{\varphi},0)$. In terms
of the function $\tau$, the Maxwell equations read
\begin{equation}
A_{\varphi,\rho}=Q\ We^\tau\tau_{,\zeta},\ \ A_{\varphi,\zeta}=-Q\
We^\tau\tau_{,\rho},
\label{elec}
\end{equation}
where $2Q^2=\gamma$. Observe that the integrability condition for the
electromagnetic function $A_\varphi$ given in (\ref{elec}) implies that
$\Delta e^\tau=0$. $\tau$ is a function which  determines alone the
electromagnetic potential. The Einstein equations written in the form
(\ref{poisson}) are convenient because we can give a direct physical
interpretation of the functions involved there. $2\lambda$ is the
gravitational potential and $e^\tau$ is the electromagnetic potential. 
If we have solved the system (\ref{poisson}), the field equation for the
$k$ function of metric (\ref{papa}) is a first order differential equation 
\begin{eqnarray}
k_{,z} &=& {1\over 4(\ln W)_{,z}}[{2W_{,zz}\over W} +
(\lambda_{,z}-\gamma\tau_{,z})^2 -  
%
%%%%% Remove the next line for preprint style:
%\cr\cr &-&
\gamma(\tau_{,z})^2\ e^{\lambda-(\gamma-2)\tau} ]    
\end{eqnarray}
where $z=\rho+i\ \zeta$. In terms of the functions $\lambda$ and $\tau$,
the thermodynamic equation reads
\begin{equation}
p_{,i}+{1\over2}(\lambda-\gamma\tau)_{,i}\ (p+\mu)=0,\ \ \ i=\rho,\ \zeta.
\label{p}
\end{equation}

System (\ref{poisson}) is a set of three linear, second order coupled
differential equations. The differential equations are coupled because the
generalized Laplace operator is determined by the function $W$, which is
determined by the pressure $p$ and the determinant of the metric $g$;
furthermore, the determinant of the metric $g$ contains all the components
of the metric. For the static Papapetrou line element (\ref{papa}) in the
harmonic map parametrization, is given by $\sqrt{-g}=W\
e^{2k-\lambda+\gamma\tau}$, which contains all the unknown function of the
system (\ref{poisson}).

In what follows we will give a general exact solution of the system
(\ref{poisson}) for the thermodynamic state equation $p=\omega\mu$. We can
set $\tau=0$ in the field equation (\ref{poisson}) and solve them and later
on solve the equation for $\tau$. We
start setting $\tau=0$. Let be $M=M(\rho,\zeta)={1\over W}$ a function
restricted to 
\begin{equation}
M (M_{,\rho\rho}+M_{,\zeta\zeta})=(M_\rho)^2 +(M_{\zeta})^2.
\label{M}
\end{equation}
In terms of $M$, an exact solution of system (\ref{poisson}) with $\tau=0$
is 
\begin{eqnarray}
e^\lambda&=&M_0M^{-d}\ e^{\lambda_1 M}\cr\cr
p&=&{M_0M^{1-d}\over k_1}\ e^{\lambda_1 M}
\label{sol}
\end{eqnarray}
where $d={3\omega+1\over 2\omega}$ and the function $k_1$ is given by
\begin{equation}
k_1=M^{-{1\over2}d^2}\exp(d\lambda_1\ M-{1\over2}\lambda_1^2\ M^2)
\label{k1}
\end{equation}
$M_0$ and $\lambda_1$ are arbitrary integration constants. 

In order to have an exact solution of (\ref{eq}), it is necessary to
fulfill the thermodynamic equation (\ref{p}). Since all the functions
involved in the solution depend explicitly on $M$, the thermodynamic
equation is a differential equation on $M$ as well. Nevertheless, all
functions involved in (\ref{p}) are already determined. Therefore equation
(\ref{p}) is a consistence equation for the two integration constants.
Equation (\ref{p}) will then determine the state equation of the perfect
fluid we are working with. We do not have solutions for arbitrary state
equations. Substituting solution (\ref{sol}) into (\ref{p}) we obtain that
$d^2=2$ and $\lambda_1=0$, this means that $\omega=-3\pm 2\ \sqrt{2}$. The
interior of this body consists of a perfect fluid which line element reads
\begin{eqnarray}
ds^2&=&{1\over f}[{M^{-{1\over 2}d^2}\over
M^2}\Omega_M(d\rho^2+d\zeta^2)+{1\over M^2}\ d\varphi^2]-f\ dt^2,\cr\cr
p&=&fM^2,\ \ \ \ \ f=M_0M^{-d},\ \ \ \Omega_M=M_{,\rho\rho}+M_{,\zeta\zeta}
\label{solp}
\end{eqnarray}
The reader can convince himself that (\ref{solp}) is an exact solution of
(\ref{eq}) and (\ref{poisson}) by direct substitution of (\ref{solp}) into
(\ref{eq}). 

The Ricci scalar of metric (\ref{solp}) is $R=-4fM^2(3-d)$ and the rest of
the invariants of this metric are powers of $fM^2$ or zero. Observe that
the norm of the time-like Killing vector ${\bf X}$ vanishes if $f=0$,
$X^{\nu}X_{\nu}=f=0$, in this region 
the pressure vanishes if $fM^2=0$. Now we have two possibilities,
$d=+\sqrt{2}$ and $d=-\sqrt{2}$.  For the first choice, $f\rightarrow 0$
implies $M\rightarrow \infty$, the Ricci scalar and the pressure are
singular for this region and all the invariants
are infinity as well. But for the second choice $d=-\sqrt{2}$ the norm of
the time-like Killing vector vanishes if $M=0$. In this region the pressure
and the invariants of this metric are regular. This second metric
represents the interior field of a body with a horizon. In this coordinates
the whole manifold needs at least two charts, one for the interior field
and at least one for the exterior field. Now the question arises; does this
metric represent a black hole? The answer depends on the exterior field we
match this metric with. Let me give an example.

For the exterior field we use the line element \cite{ma25}
\begin{eqnarray}
ds^2&=&{1\over f}[{N^{-{1\over2}d^2}\over N^3}\Omega_N(d\rho^2+
d\zeta^2)+{1\over N^2}\ d\varphi^2]-f\ dt^2,\cr\cr
f&=&N_0N^{-d}
\label{solv}
\end{eqnarray}
where now $p=\mu=0$. It is easy to see that the metric (\ref{solv}) is an
exact solution of the vacuum Einstein equations for arbitrary constants
$N_0$ and $d$, provided that
\begin{equation}
N (N_{,\rho\rho}+N_{,\zeta\zeta})=2((N_\rho)^2 +(N_{\zeta})^2).
\label{N}
\end{equation}
Compare the equations (\ref{M}) and (\ref{N}), the first one is for fields
inside of the object and the second one is for outside of it. One solution
of (\ref{M}) is $M={1\over \rho^2+\zeta^2}$ and one solution of (\ref{N})
is $N={1\over \sqrt{2}\rho}$. For these solutions the line elements now
read
\begin{eqnarray}
ds^2&=&{1\over f}[4\,M^{-1}(d\rho^2+d\zeta^2)+{1\over M^2}\ d\varphi^2]-f\
dt^2,\cr\cr
p&=&fM^2,\ \ \ \ \ f=M_0M^{\sqrt{2}},
\end{eqnarray}
for inside of the body and
\begin{eqnarray}
ds^2&=&{1\over f}[4\,N^{-1}(d\rho^2+d\zeta^2)+{1\over N^2}\ d\varphi^2]-f\
dt^2,\cr\cr
f&=&N_0N^{\sqrt{2}}
\label{solvt}
\end{eqnarray}
for outside of it.
It is now easy to match the solutions, we need only to choose $M|_R=N|_R$,
where $R$ is the boundary of the object. In what follows I investigate this
region.  In these coordinates $\rho$ and $\zeta$ can take all values inside
of the object provided that 
$\rho^2+\zeta^2\not=0$. The region $R$ corresponds to
\begin{equation}
\frac{1}{2}{\frac {\sqrt {2}\left (-\sqrt {2}\rho+{\rho}^{2}+{\zeta}^{2}
\right )}{\left (\rho^2+\zeta^2\right)\rho}}=0
\label{S}
\end{equation}
which has three solutions, two solutions of (\ref{S}) are
$\rho|_R=\frac{1}{\sqrt {2}}\pm 1/2\,\sqrt {2-4\,{\zeta}^{2}}$ and the
third solution corresponds to $\rho|_R\rightarrow\infty$. Inside of the
object, the pressure reads 
\[
p=\frac{M_0}{(\rho^2+\zeta^2)^{2+\sqrt{2}}}.
\]
For the two first solutions of (\ref{S}) the pressure is not necessarily
small or zero. Therefore the boundary of the object corresponds to
$\rho|_R\rightarrow\infty$, where $p\rightarrow 0$. A surface of constant
pressure is
\[
\rho^2+\zeta^2=(\frac{M_0}{p})^{\frac{1}{2+\sqrt{2}}}=r_0^2, 
\]
which represents a sphere of radios $r_0$ in the plane
$(\rho,\zeta,\varphi)$. Strictly speaking the pressure will be zero at
$\rho>>1$, but in fact this surface in not very big. Suppose that
$p=10^{-6}M_0$, which could be a very small pressure, in this case the
boundary surface has a radios of $r_0=7.5627$ which is not too big.
Therefore the boundary surface of this object is a sphere with $p<<M_0$ and
radios $r_0=(\frac{M_0}{p})^{\frac{1}{2(2+\sqrt{2})}}$.

In Boyer-Lindquist coordinates 
\[
\rho=\sqrt{r^2-2mr+\sigma^2}\sin\theta,
\] 
\[
\zeta=(r-m)\cos\theta,
\] 
the metrics read
\begin{equation}
ds^2={1\over f}[{1\over4}k_1\Omega_M({dr^2\over r^2-2mr+\sigma^2}+
d\theta^2)+W^2\ d\varphi^2]-f\ dt^2,
\end{equation}
where 
\[
\Omega_M=M((\sqrt{r^2-2mr+\sigma^2}\
M_{,r})_{,r}\sqrt{r^2-2mr+\sigma^2}+M_{,\theta\theta})
\]
and the equations (\ref{M}) and (\ref{N}) transform into
\[
M((\sqrt{r^2-2mr+\sigma^2}\
M_{,r})_{,r}\sqrt{r^2-2mr+\sigma^2}+M_{,\theta\theta})=
\]
\begin{equation}
n((M_{,r})^2(r^2-2mr+\sigma^2)+(M_{,\theta})^2).
\label{Mbl}
\end{equation}
where $n=1$ corresponds to (\ref{M}) and $n=2$ to (\ref{N}).
Using separation of variables one finds that the solutions of equation
(\ref{Mbl}) are given in terms of hypergeometric functions, but this is
material for further investigations \cite{fco}

Now let me magnetize this body, $i.e.$, now $\tau\not=0$. We must solve the
Maxwell equation in (\ref{poisson}) for this particular value of $\lambda$.
Nevertheless, the equation for $\tau$ in (\ref{poisson}) should be in
agreement with the integrability 
conditions for the electromagnetic potential $A_\varphi$ in (\ref{elec}).
But the integrability conditions for $A_\varphi$ are
\begin{equation}
\Delta e^{\tau}=\Delta\tau+2M((\tau_{\rho})^2+(\tau_{\zeta})^2)
\end{equation}
which seems impossible to be fulfilled because of the equation for $\tau$
in (\ref{poisson}), unless $\lambda-(\gamma-2)\tau=0$. This seems not to be
the case because $\lambda$ and $\tau$ fulfill very different differential
equations. But if we choose $\gamma=2+d$ it happens a miracle. Using the
solution (\ref{sol}) we observe that $\tau={1\over
\gamma-2}\lambda=\ln({1\over M_0M})$ which implies that $\Delta
e^{\tau}=0$. The magnetized metric now reads
\begin{eqnarray}
ds^2&=&{1\over f}[M^d\Omega_M(d\rho^2+d\zeta^2)+M^2d\varphi^2]-f\ dt^2\cr\cr
f&=&M_1M^2\cr\cr
p&=&{e^{\lambda-\gamma\tau}\over M\ k_1}=M_1M^{1-d}\cr\cr
\tau&=&-{1\over2}\ln(2f)
\end{eqnarray}
where $M$ fulfills (\ref{M}).
The thermodynamic equation (\ref{p}) fixes again the constants. In this
case we obtain $d=1$, that means $\omega=-1$ and $Q^2={1\over 2M_1}$. For
this values of the constants the pressure becomes constant in all of
space-time and the invariants are powers
of $-8M_1\ \ i.e.$, constants, therefore this metric is regular all over.
But now the pressure cannot vanish at all. This object is a magnetized
perfect fluid with state equation $p=-\mu$, which covers the whole
space-time with a constant pressure.

To obtain the space-time of a more realistic object it is necessary to
study other state equations \cite{fco}. Using equations (\ref{poisson}) and
the procedure for integrating them given here, it is possible to
investigate the interior behavior of these 
more realistic bodies \cite{fco}.

I want to thank the relativity group in Jena for his kind hospitality. This
work is partially supported by CONACYT-M\'exico, grant 3697-E.


\begin{thebibliography}{99}

\bibitem{kra} D. Kramer, H. Stephani, M. MacCallum and E. Held. {\it Exact
Solutions of Einstein Field Equations.} (1980), DVW, Berlin.

\bibitem{ma25} T. Matos. {\it Phys. Rev.} {\bf D49}, 4296, (1994).

\bibitem{ma36} T. Matos, D. Nu\~nez and H. Quevedo. {\it Phys. Rev.} {\bf
D51}, R310, (1995).

\bibitem{fco} T. Matos, F. Sidharta and H. Villegas. In preparation.




\end{thebibliography}
\end{document}